# Feasibility of Utilizing the CTA Gamma-Ray Telescopes as Free-Space Optical Communication Ground Stations

Alberto Carrasco-Casado,[1*] Mariafernanda Vilera,[1] Ricardo Vergaz,[1] and Juan F. Cabrero[2]

[1] *Universidad Carlos III de Madrid, Avda. Universidad, 200, Madrid, 28911, Spain*
[2] *INSA Aerospace Services and Engineering, Paseo del Pintor Rosales, 34, Madrid, 28008, Spain*
*Corresponding author: aacarras@ing.uc3m.es*

The signals that will be received on Earth from deep-space probes in future implementations of free-space optical communication will be extremely weak and new ground stations will have to be developed in order to support these links. This paper addresses the feasibility of using the technology developed in the gamma-ray telescopes that will make up the CTA (Cherenkov Telescope Array) observatory in the implementation of a new kind of ground station. Within the main advantages that these telescopes provide are the much larger apertures needed to overcome the power limitation that ground-based gamma-ray astronomy and optical communication both have. Also, the large number of big telescopes that will be built for CTA will make it possible to reduce costs by scale-economy production, enabling optical communications in the big telescopes that will be needed for future deep-space links.

## 1. INTRODUCTION

There is more and more consensus on the utilization of optical communication in deep-space communication links as the distance and data-rate requirements increase [1-3]. The orders-of-magnitude-lower divergence of optical wavelengths compared to radio frequency allow a much more efficient power delivery from the transmitter to the receiver, and a great deal of optical fiber technology can be used in free-space links, providing much faster communication rates than RF, among other advantages such as lower payload power, size and weight [4]. However, the future optical signals received on Earth from deep-space probes will be extremely weak, demanding the most from each element of the chain. Ground stations hold great potential to improve the reception of these signals [5], the most straightforward single measure being the larger aperture needed to capture more photons (the same measure has been taken in RF: being the 70-meter antennas used as the main receivers of NASA's Deep Space Network).

The Cherenkov Telescope Array (CTA) project, currently in its design phase, is a multi-national collaboration to build within the next few years a new generation of Imaging Atmospheric Cherenkov Telescopes (IACT) based on the success of ground-based gamma-ray astronomy in recent years [6]. CTA will achieve an order-of-magnitude sensitivity improvement by deploying an array of several tens of telescopes of three different sizes (~6 m, ~12 m and ~24 m in diameter) in two different locations –one in each hemisphere– yet to be determined. Gamma-ray astronomy shares the same limitation with deep-space lasercom, i.e., extremely weak signals to be detected at the ground site. The solution taken in Cherenkov telescopes has been traditionally, and will be in CTA as well, to increase the reception apertures and to replicate them by using array topologies. These solutions are also shared with free-space optical communications (FSOCs), although the magnitude of the CTA telescopes apertures hasn't been even proposed so far. The main reason is the delay in the deployment of lasercom technology in deep-space scenarios, along with the fact that the largest apertures needed to support first attempts at deep-space links are below half the largest ones designed for CTA.

IACTs don't detect gamma-ray radiation directly. Instead, they detect the effects of this radiation after interacting with the atmosphere. When particles that result from this interaction of high-energy gamma-rays with molecules in the upper atmosphere travel through a medium faster than the speed of light in that medium, they induce a Cherenkov effect, consisting of a cone-shaped shower of photons directed along the path determined by the original particles [7]. The Cherenkov photons' wavelengths range in wavelength from 300 nm up to several meters, limited by the atmospheric transmission [8]. However, the spectral intensity distribution is proportional to $\lambda^{-2}$ and thus ultraviolet and blue components predominate. For this reason, although the goal of CTA is to perform gamma-ray astronomy, IACTs are indeed optical telescopes and a number of favorable circumstances justify studying the possibility of taking advantage of the CTA technology to develop FSOC ground stations.

In the following sections, these motivations along with the considerations involving the utilization of CTA's Cherenkov telescopes as FSOC ground stations will be taken into account, emphasizing the ones that imply some kind of adaptation or modification from the original telescope design in order to make it suitable to work as a communication terminal. Given the current state of CTA project, in which there is no final design for the telescopes yet, whenever possible, information from CTA project has been used. In the cases that the design is not finished or the data is not available, information from one of the currently most advanced Cherenkov telescopes under operation has been applied. These telescopes are MAGIC I & II (Major Atmospheric Gamma-ray Imaging Cherenkov) in the Roque de los Muchachos observatory at La Palma [9].

## 2. MOTIVATIONS

In this section, a summary review is presented with the main reasons that justify the study of the utilization of CTA telescopes as FSOC ground stations.

1. Deep-space FSOCs are based on photon-starved links by nature, and this limitation increases with the square of the distance. A major link margin improvement is to increase the receiver aperture and CTA telescopes provide larger apertures than the largest apertures ever considered for optical ground stations.
2. CTA development phase will be of such magnitude that it can be compared to a telescope assembly line, in which the costs of producing additional units are lower than a single development. The costs of the infrastructure to support the facilities are also lower than a dedicated one due to the shared location and operation.
3. IACTs operate in array topology by nature, which could be taken advantage of for increasing arbitrarily the effective receiver aperture by replicating elements if needed.
4. The optical quality of IACT mirrors is much less demanding than in astronomical telescopes, which implies much lower costs. This is in principle a shared feature with FSOC, although as will be explained, some corrections are needed to allow the communication link to work in worst-case scenarios.
5. FSOC and Cherenkov observation differ in the spectrum region to detect, so the wavelength-dependent mirror reflectivity is a limiting parameter. As will be explained, the reflectivity happens to be even more advantageous at communication wavelengths, allowing mirror reutilization without either modification or reflection losses.
6. The fast nature of Cherenkov signals demands an electronics and communications infrastructure that satisfies FSOC needs. GHz-sampling of cameras with hundreds of pixels each and Tbit/s optical-fiber transmission from the telescopes provide a capacity currently more than enough to support the communication signals.
7. Earth-based gamma-ray astronomy locations share the same requirements as lasercom ground stations regarding atmospheric conditions: high altitude to avoid turbulence, absence of luminous sources and low impact from clouding and scattering.
8. CTA foresees two locations –one in each hemisphere– equidistantly separated in longitude, which would be an added value regarding the blockage effect of a rotating Earth in deep-space missions.
9. One of the main goals of CTA is gamma-ray burst detection, which demands a very fast telescope repositioning. This feature enables near Earth lasercom with high-speed satellites such as LEO.
10. IACT's reflector shape is chosen according to two requisites: minimize temporal dispersion and coma aberration. The first one is shared with FSOC due to the similar speed of Cherenkov events and communication signals, and the second one is more demanding in IACTs due to the uncertainty in the direction of the signals.

## 3. DISCUSSION

### 1. IACT mirrors

As was introduced, optical components predominate in the Cherenkov radiation that reaches the ground, with a maximum at about 350 nm. Therefore, telescope mirrors are designed in order to maximize the reflection around this region. Although FSOC can take place around the Cherenkov spectrum [10], the most probable wavelength to be implanted in this kind of link in the future is 1550 nm because of multiple reasons, such as the amount of technology developed for the optical fiber 3rd window and less-adverse effects in atmospheric propagation compared to shorter wavelengths. Most relevant space agencies agree on this [11], hence in this paper only this wavelength has been taken into account. Since 1550 nm is far from Cherenkov radiation maximum, there are no measurements of mirrors in this region of the spectrum. The authors of this paper gained access to MAGIC telescope facilities in La Palma, Canary Islands (Spain), and took reflectivity measurements directly from one of its mirrors.

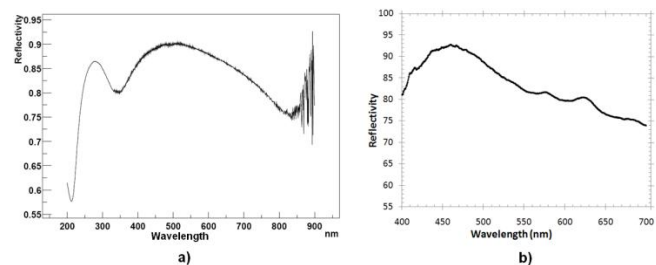

Fig. 1. Reflectivity measurement of a MAGIC mirror in the visible spectrum according to [12] (a) and to the authors of this paper (b).

These measurements were taken using two portable spectrometers, one for 200-1100 nm and other for 900-1700 nm. The first one, covering the visible spectrum, was used in order to compare the measurements with previous ones [12]. A comparison between them (Fig. 1) shows a good agreement in the band of interest for Cherenkov astronomy, 400-600 nm, where the reflectivity is always over 80% in both cases. Little deviations observed are due to coating variations, which can exceed 5%, depending on the region of the mirror [8]. In Fig. 2, reflectivity measurements in the infrared band are presented, taken over different sections of the mirror.

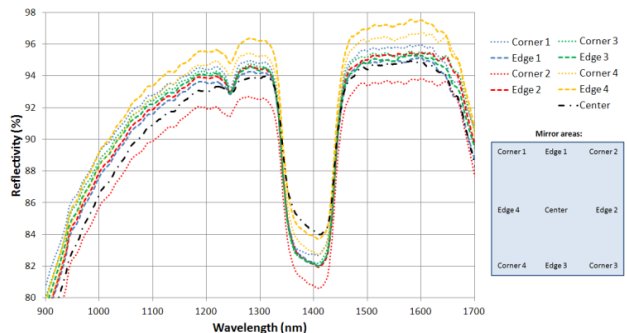

Fig. 2. Reflectivity measurements of a MAGIC mirror in the IR band taken over different sections

In Fig. 3, the average reflectivity measurements taken from 400 nm to 1700 nm are shown. Although the MAGIC mirrors were not designed aiming to optimize the reflectivity in the IR region, this is exactly what happens, especially at the communications wavelength, presenting a maximum around 1550 nm, where the reflectivity reaches 95%. It can be concluded that IACT mirrors can be directly used in FSOC, even with a better performance than in Cherenkov astronomy itself and with very little difference with conventional astronomy telescopes: only 0,2 dB compared to the one based on a multi-layer dielectric interference coating with 99.7% reflectivity [13].

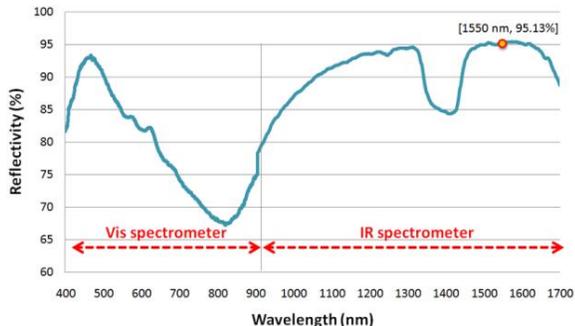

Fig. 3. Average reflectivity measurements of a MAGIC mirror

### 2. Focusing capability and detector

The focusing potential of a telescope determines its capability to concentrate the light gathered by the main reflector coming from a source at the infinity and making it converge at one point. In both FSOC and conventional astronomy the source is so far away that it can be considered to be at infinity, hence the detector or camera is positioned at the focal length. IACT's approach is different because the light to be detected comes not from the infinity, but from an unknown distance between 6 and 20 km high in the atmosphere [14]. These telescopes are usually focused considering the source at 10 km, so any other distance is defocused, including the infinity, and the camera is placed at a different distance from the focal length. MAGIC telescope has a 17 m reflector with a f/D ratio of 1.03 [15], therefore the focal length $f$ is 17.51 m. Applying the image formation equation [16]

$$\frac{1}{f}=\frac{1}{s}+\frac{1}{s'}=\frac{1}{s_\infty}+\frac{1}{s'_\infty}=\frac{1}{10km}+\frac{1}{s'_\infty+\varepsilon} \quad (1)$$

where $s$ is the distance from the object and $s'$ the distance to the image, so if $s_\infty = \infty$, then $s'_\infty = f$ and using $s=10$ km, it is possible to find the displacement $\varepsilon$ from the original camera position that has to be applied in order to focus. This displacement is 3 cm, which is within the range that the camera is allowed to move in order to do the original calibration [17], so although in principle IACTs are not designed to focus at infinity, this can be achieved easily by a slight movement of the camera.

Regarding the displacement discussed, it is important to note another big difference between a communication and a Cherenkov telescope: while in the first one the photosensitive device is a single detector, IACTs use "full" cameras. These cameras are made up of thousands of single detectors –usually photomultipliers (PMT)– and weigh up to 2.5 tons [18], so such a movement is a permanent modification. Since the receiver, including the photodetector and the signal processing, is a much smaller system than the Cherenkov camera, the adaptation is certainly feasible. The original detectors cannot be reutilized due to their low quantum efficiency in 1550 nm, –although their time response is adequate, as the Cherenkov events occur at a similar speed as communications–, so they have to be replaced with a FSOC detector. This system would not differ from the ones that would be used in an equivalent deep-space optical ground station, so it won't be discussed here.

### 3. Field of view and background noise

The field of view (FOV) of a telescope is a fundamental feature to assess the signal-to-noise ratio (SNR) in a FSOC link. The relation between the FOV and the SNR is determined by the background noise if the FOV is much bigger than the source, so the tendency should be to decrease the field of view as much as possible [19]. In FSOC, the remote terminal is equivalent to a point source at infinity; hence, the FOV could be near zero, ideally. In practical terms, there are a number of limitations that move it away from zero. The first of them is the diffraction limit $\theta=2.44\lambda/D$ [20]. In a Cherenkov telescope, a FOV close to this limit could never be used, even if the optics quality could achieve it, which is not the case. For example, a CTA-LST (Large-Size Telescope), with ~24 m in diameter, working in 1550 nm, would impose a diffraction limit of 0.0000045°, an unachievable FOV, as will be explained below.

Table 1. Security margin as a function of pointing and field of view of several telescopes in operation

| Telescope | Type | Pointing resolution | Field of view | Security margin |
|---|---|---|---|---|
| MAGIC | IACT | 0.016° [27] | 0.1° [16] | 6.25 |
| VERITAS | IACT | 0.01° [23] | 0.11° [23] | 11 |
| CTA-LST | IACT | 0.003° [29] | 0.07°-0.12° [30] | 23.33-40 |
| OCTL | FSOC | 0.00083° [31] | 0.028° [31] | 34.5 |
| OLSG | FSOC | 0.00027° [11] | 0.00138° [11] | 5 |
| 10mOGS | FSOC | 0.0057° [32] | 0.0057° [32] | 1 |

There is another limiting factor above the diffraction limit: the pointing resolution. The FOV has to be at least equal to the pointing resolution to assure that the target is always seen by the telescope, although usually a larger security margin is chosen (Table 1). The design of the FOV in FSOC is in fact determined by this limit [21], rather than the diffraction limit. In Cherenkov and communication telescopes, a comparable ratio between field of view and pointing resolution is used. However, the FOV in FSOC telescopes is much narrower. In part, this is explained by the more demanding pointing of conventional telescopes compared to IACTs −which allows to reduce the field of view. Although the pointing performance has been improving, there is still an important gap between them.

Nevertheless, the main reason explaining the FOV difference lays in another limit above pointing resolution: the optical quality of the mirrors, characterized by the point spread function (PSF). The PSF measures the reflector capability to concentrate the light from a point source at the focal plane. In practice, it defines a bell-shaped blurred area, usually fitted with a Gaussian function. It determines the angular resolution of the telescope and it is the most important difference between Cherenkov and optical telescopes, with a gap of up to 3 orders of magnitude between them [22]. The reason for this is related to the camera's FOV and its number of pixels: IACTs must observe a wide fraction of the sky due to the intrinsic size of Cherenkov showers and the uncertainty of their origin. A wide FOV and a high angular resolution are commonly incompatible unless the pixels can be miniaturized in order to have a high number of them, which is not the case with IACT cameras, currently.

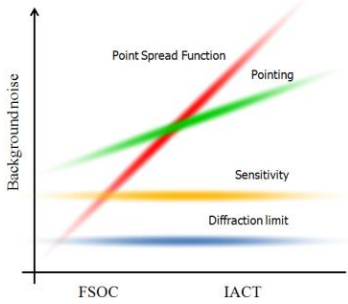

Fig. 4. Background noise limitation in IACTs and FSOC telescopes

In FSOC, the FOV should be bigger than the PSF and this is the worst limit imposed by an IACT in terms of background noise. Therefore, it can be concluded that the PSF is what in fact limits the performance of Cherenkov telescopes. This fundamental conclusion can be confirmed this way: assuming an optimistic security margin of 5 and a PSF of 0.02º in CTA-MST (Medium-Size Telescope) [29], the minimum FOV should be 0.1º, which is a much bigger FOV than in FSOC telescopes. For this reason, in order to increase the SNR of the communication link, the efforts should be focused on improving the PSF of IACTs. Fig. 4 summarizes the limits related to background noise that has been discussed and it shows qualitatively where every limitation would appear when completing each telescope upgrade.

### 4. Angular resolution improvement

In a FSOC link, a wide FOV like the one studied in the previous section could make communication impossible in presence of high levels of background noise, namely, during the day. For this reason, the PSF should be improved in CTA telescopes to make them suitable to lasercom. This section deals with proposals to accomplish this goal.

The main cause behind the poor PSF in IACTs is directly related to the optical quality of the reflector. These reflectors are made up of an array of mirror segments forming a predefined shape. In big Cherenkov telescopes, the shape usually employed is parabolic –due to its isochronous property, which allows photons to reach the focus at the same time coming from different parts of the surface [24]. The parabolic shape is achieved by using a different radius of curvature in each segment and by placing them in their corresponding position according to the paraboloid. A simulation of this parabolic-equivalent shape has been performed using the ray-tracing software *OpticsLab* by *Science Lab Software* in order to evaluate the PSF and be able to assess the result of the proposals. To validate these results, the MAGIC-II telescope reflector has been used in the simulations to be able to make a later comparison between the simulation results and experimental measurements.

MAGIC-II has a reflector surface of 247 m² made up of 249 square mirrors of 98.5 cm side each. The parabolic shape is defined by its f/D=1.03 with D=17 m and f=17.51 m, as was stated in section 2.2. Each mirror was placed according to the equation of a 2D paraboloid centered in {0,0} with the focus in {17.51, 0} and of a shape given by $y^2=4fx=70.04x$, distributing the mirror segments equidistantly along this curve. Mirrors are spherical with different radius of curvature in each mirror depending on their distance to the center of the parabola and they were obtained using the average ($rc_{ave}$) between the maximum ($rc_{max}$) and the minimum ($rc_{min}$) value in each segment. The simulation used 61,517 rays and resulted in a PSF of 2.61 cm in the FOV center, which translates into an angular extent of 0.085º, according to $\theta_{FOV}=2atan(PSF/2f)$. This figure is consistent with an experimental measurement, according to which the 2.61 cm would result in 79% encircled energy [25] considering that the PSF is a Gaussian distribution with σ=10.5 mm, as measured in [26]. The agreement between the simulated result and the experimental measurement establishes the validity of this simulation to at least qualitatively assess the proposals for improving the PSF. In this way, it is possible to check the origin of the poor angular resolution compared to the PSF of an ideal parabolic-shape version (Fig. 5.a) of the MAGIC-II telescope and its segmented version (Fig. 5.b).

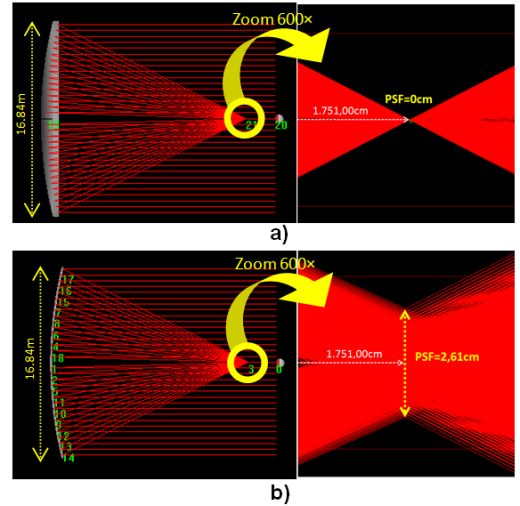

Fig. 5. 33-ray simulation and 61,517-ray PSF estimation in MAGIC-II single-mirror (a) and segmented-mirror (b)

The two improvements proposed next intend to bring the final shape of the reflector closer to the paraboloid, since this has been identified as the best strategy to reduce the PSF. The first one consists in subdividing each mirror facet into a set of four or more smaller segments, all mounted on the same panel. By this way, the structural change is minimum and there is no need to change the mirror polishing technique, since the radius of curvature is fixed, yet the final shape is closer to the ideal paraboloid as there are more different radii. This approach has been simulated dividing each MAGIC 1 m² mirror in four 0.5 m² segments, with different but constant radii of curvature, corresponding with their position in the paraboloid. If each mirror is accurately focused –only a reorientation of less than 0.4º is needed– the range of the PSF can be reduced to less than 1 cm. The second proposal consists in polishing each mirror to make the radius of curvature variable depending on the surface location, so that each of them is a little part of the final parabola. This way, there is no structural change and all the efforts focus on a more accurate polishing, similar to the ones employed in astronomical telescopes, and these techniques are well proven. It is important to note that at 1550 nm the accuracy of the polishing is less demanding than in visible wavelengths.

### 5. Detection size optimization

If the PSF could be improved at least up to the point that it was the pointing resolution what limited the FOV, then the FSOC performance of IACTs could be comparable to conventional telescopes and all the advantages of Cherenkov telescopes, especially larger apertures and smaller costs, would manifest. Nevertheless, in this section another approach is proposed in order to deal with the poor PSF, assuming that such improvement is not carried out. In any telescope, the detection area in the focal plane determines the field of view [23]. In an IACT, the large PSF imposes a large detection area if the entire signal has to be collected. However, if the detection area is reduced, the FOV is also reduced and so the background noise. The drawback is that less power reaches the detector.

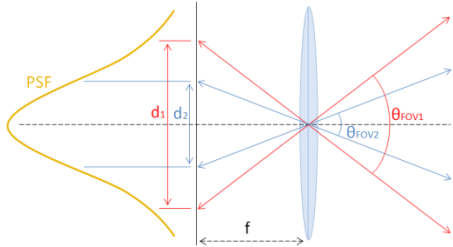

Fig. 6. Relation between detection area and field of view

In Fig. 6, it can be seen that a detection size d1 is enough to collect most of the signal, determining a field of view $\theta_{FOV1}$. However, if the detection size is reduced, the FOV is also reduced, which has two implications: on one hand, the background noise integrated by the detector is reduced, and on the other hand the signal power is also reduced. The point is which of them has been reduced more. In Fig. 7, the result of a link budget as a function of the detection size is shown. A worst-case framework was used in the calculation, consisting in a Lagrange point L1 scenario in which the link is always established during the day. Nevertheless, the parameters of the simulation are not particularly relevant, since the key point to study here is the trend of SNR, which is always the same in presence of daylight. It can be established that when detection size is decreased, background noise decreases faster than signal power, thus the SNR will be maximized if the detection area is minimized. However, an unlimited reduction would end in no photons reaching the photodetector, so the detection size can be reduced only up to the receiver sensitivity, also shown in Fig. 7. The sensitivity is always well below the received signal but cannot be achieved –so the FOV cannot be reduced up to that point– because the pointing resolution is above the sensitivity, and that will be the parameter that effectively limits the minimum FOV.

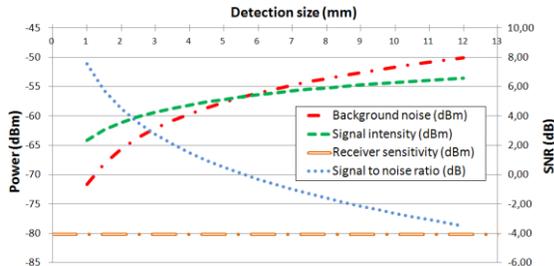

Fig. 7. Background noise power, signal power, sensibility and SNR as a function of the detection size

### 6. Pointing and tracking

Pointing capability in Cherenkov telescopes does not present any special requirement regarding resolution compared to conventional telescopes, therefore commercial components are usually employed [27]. Pointing resolution mainly relies on shaft encoders, that are used to accurately determine the information of instantaneous position of azimut/elevation gears and close the feedback loop with the target position. IACT's maximum resolution is limited by the nominal shaft encoder resolution, which determines the number of different positions that can be encoded. Based on this specification, it is possible to compare (Fig. 8) the maximum resolution of several telescopes, obtaining the pointing resolution as $360°/2^N$, with N as the encoder resolution in bits. It is clear that IACTs pointing is worse than in conventional telescopes, which can be explained through their poor optical resolution, which makes better pointing useless, so improvements are indeed possible in order to reduce the FOV.

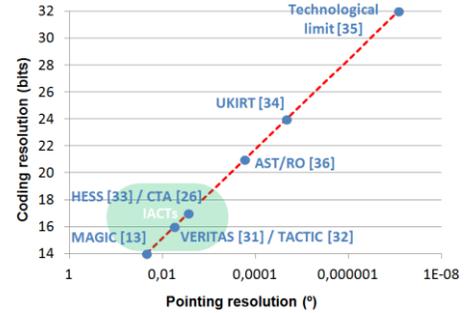

Fig. 8. Encoder resolution vs pointing resolution for several telescopes

Regarding tracking capability, the stress lies now in the speed at which the telescope can follow a moving target, so the requirements are related to fast sources such as Low-Earth orbiting (LEO) satellites. Although the main goal of a ground station based on IACTs is deep-space probes, a multipurpose receiving terminal that could track all kinds of targets would add value to the network in which it worked. Fortunately, one of the main objectives of CTA is the study of gamma-ray bursts, for which the telescopes have to be reoriented very fast after a satellite warning. MAGIC can point to any point in the sky in 30 s maximum [28], and CTA telescopes will move as fast as 180° in 20 s [29], or 9°/s. Considering 160 km as the minimum LEO orbital altitude $R_{LEO}$, according to eq. 2 [33], $R_T$ being the Earth radius, $G$ the gravitational constant and $M$ the mass of the Earth, an orbital period $T$ of 5,261.29 s is obtained, which results in an angular speed of 0.068°/s, 131 times lower than CTA tracking capability, therefore this is not a limitation.

$$T = 2\pi\sqrt{\frac{r^3}{G \cdot M}} = 2\pi\sqrt{\frac{(R_{LEO} + R_T)^3}{G \cdot M}} \qquad (2)$$

### 7. Daylight and shared operation

Cherenkov telescopes are only active during the night, never in presence of daylight. Even during the night, there cannot be any strong light source, the Moon being the most important one, e.g. MAGIC telescope was specifically designed in order to tolerate moderate Moonlight conditions [40]. An FSOC ground station should work during day and night, bearing high levels of daylight and even pointing as near to the sun as a few degrees [41]. However, IACT's technical incapability regarding background light is only related to the high sensitive PMTs, and as was explained, these photodetectors should have to be replaced with the ones usually employed in FSOC receivers. Nevertheless, the fact that Cherenkov telescopes are only operated during night, forces us to adapt them to the presence of daylight. These adaptations aren't treated here, as they have been extensively studied in other works [42] due to the increasing interest on reutilizing existing astronomical facilities for optical communications. Although plenty of solutions are available, the case for daylight adaptation of IACTs present several singularities –some of them favorable, such as their total exposure to the weather design, and some of them adverse, such as the bigger size of the telescopes–, so their outdoor adapted design should be studied independently. Protecting the detector, maintaining the mirrors shapes under their tolerances and take into account atmospheric turbulence are some of the matters to be considered in future works.

In this paper, the case for reutilization of IACTs has been focused on adapting a telescope for FSOC operation exclusively. However, there exists the possibility of sharing the same telescope for astronomy and lasercom in order to make the most of it as there are times in which it cannot be used for observation, namely, during the day or with moonlight. The most important adaptations have their origin in the different spectral sensitivity and the different focusing requirements. The first one makes it necessary to use different detectors for each purpose, as Cherenkov light and communication signals are in separate spectral regions. Fortunately, the FOV requirements of each operation allow using compatible detectors: gamma-ray astronomy demands a wide field of view, which needs a big camera, and lasercom demands a narrow field of view, which needs a small photodetector. This way, it is certainly possible to replace one of the pixels of the Cherenkov camera with the FSOC detector. A similar approach was carried out in 2005 when the central pixel of MAGIC camera was replaced with a detector dedicated to visible astronomy [43]. The PMTs used in IACTs are extremely sensitive, up to the point that they could be damaged if exposed to daylight, so a protection mask should cover the PMTs excluding the communication pixel, blocking the light to the Cherenkov Camera and letting it pass to the FSOC detector.

As was stated, the second big difference between Cherenkov operation and FSOC operation refers to the focusing requirements: in section 3.2, the need to move the camera 3 cm away from the focus was explained. This reallocation was justified by means of focusing at infinity, instead of at 10 km. If the telescope has to be used for both astronomy and lasercom, both focusing capabilities have to be dealt with, and moving the camera is not a viable solution, as it involves a manual calibration each time, –although a simple solution could be to move only the communication detector, if this was physically possible. A more suitable solution is proposed here involving the use of the Active Mirror Control (AMC). AMC is a system that allows the readjustment of each mirror panel to a pre-known position in order to correct in real time the deformations that take place in the reflector structure due to variations of mass distribution with the different pointing direction [44]. By using this system it is possible to change the paraboloid shape enough to focus at a different focal point, so if the camera is placed in its original position –at 17.54 m from the reflector–, the reflector could be reshaped to focus at infinity as if the focal point was 17.54 m and not 17.51 m as it actually is. A simulation has been performed to validate this proposal using the MAGIC paraboloid that was modeled in section 3.4. The mirror segments that made up the reflector were adjusted to focus a point source at infinity in 17.54 m and the PSF was measured. This new PSF resulted in 2.67 cm, which is less than half the PSF in the same position before the realignment, and a very close result to the original PSF moving the camera to 17.51 m –only 2.25% bigger than this 2.61-cm PSF. This proposal allows switching between Cherenkov mode and lasercom mode almost instantly and doesn't interfere with the normal operation of AMC correction since the maximum movement of a mirror is 0.02°, well below the AMC dynamic range.

## 4. CONCLUSIONS

In this paper, a proposal has been made to reutilize the technology developed in the gamma-ray telescopes of the CTA project for the implementation of enhanced optical ground stations to support missions that could span from LEO to deep-space and could extend the range of distance and performance of free-space optical communications.

The reasons that justify this study are varied and the most relevant include a cost reduction in the development of telescopes with very large apertures, which provides the highest receiver gains ever considered in a FSOC link. Besides, CTA facilities share a number of features with lasercom such as favorable atmospheric conditions or fast electronics and communications infrastructure, which allows a natural integration of FSOC terminals within the CTA facilities.

A number of adaptations and upgrades are needed though, due to differences in the operation of each kind of telescope. In this paper, the most relevant ones that relate to a laser communication receiver have been studied, as well as solutions proposed and evaluated. The most influential enhancement is related to the angular resolution in order to reduce the FOV and the background noise. To achieve this, several approaches have been suggested, including a PSF improvement and a detection size optimization. Other operational differences have been studied, including mirror reflectivity, focusing capability and pointing.

In this study no major limitations have been found that disable Cherenkov telescopes to operate as FSOC ground stations after the required modifications. Since to our knowledge this is the first time this proposal has been suggested, further research is needed on how to implement each adaptation in detail and how these terminals would behave in a real communication link, which is being currently carried out by the authors of this paper.

## ACKNOWLEDGEMENTS

We thank the support by Comunidad Autónoma de Madrid (grant S2009/ESP-1781, FACTOTEM-2), by CDTI Centro para el Desarrollo Tecnológico Industrial (grant IDC-20101048 awarded to the company INSA within the Industry of Science Plan) and the helpful assistance by José Luis Contreras, from High Energy Physics Group, Universidad Complutense de Madrid.